\documentclass[twocolumn]{aastex701}
\shorttitle{Detecting Circumplanetary Rings with Direct Imaging}
\shortauthors{Hasler et al.}
\graphicspath{{./}{figures/}}

\newcommand{\blue}[1]{\textcolor{black}{#1}}
\usepackage{siunitx}

\begin{document}

\title{Entering the Era of Circumplanetary Ring Detection with Direct Imaging}

\correspondingauthor{S.~N.~Hasler}
\email{shasler@stsci.edu}

\author[0000-0002-4894-193X]{Samantha N.~Hasler}
\affiliation{Space Telescope Science Institute,
3700 San Martin Drive,
Baltimore, MD 21218, USA}
\affiliation{Department of Earth, Atmospheric, and Planetary Sciences, Massachusetts Institute of Technology, 
77 Massachusetts Avenue, Cambridge, MA 02139}
\affiliation{IPAC, California Institute of Technology,
1200 E. California Boulevard,
Pasadena, CA 91125, USA}
\email[]{shasler@stsci.edu}

\author[0000-0002-7162-8036]{Alexandra Z.~Greenbaum}
\affiliation{IPAC, California Institute of Technology,
1200 E. California Boulevard,
Pasadena, CA 91125, USA}
\email{azg@ipac.caltech.edu}

\author[0000-0001-5966-837X]{Geoffrey Bryden}
\affiliation{Jet Propulsion Laboratory, 
California Institute of Technology, 
4800 Oak Grove Drive, Pasadena, CA 91109, USA}
\email{geoffrey.bryden@jpl.nasa.gov}

\author[0000-0002-5407-2806]{Vanessa Bailey}
\affiliation{Jet Propulsion Laboratory, 
California Institute of Technology, 
4800 Oak Grove Drive, 
Pasadena, CA 91109, USA}
\email{vanessa.bailey@jpl.nasa.gov}

\author[0000-0002-3414-784X]{Jorge Llop-Sayson}
\affiliation{Jet Propulsion Laboratory, 
California Institute of Technology, 
4800 Oak Grove Drive, Pasadena, CA 91109, USA}
\email{jorge.llop.sayson@jpl.nasa.gov}

\author[0009-0007-5843-6447]{Elizabeth Lane}
\affiliation{Department of Astronomy, 
University of Michigan, 
Ann Arbor, MI 48109, USA}
\email{laneeli@umich.edu}

\author[0000-0002-9521-9798]{Mary Anne Limbach}
\affiliation{Department of Astronomy, 
University of Michigan, 
Ann Arbor, MI 48109, USA}
\email{mlimbach@umich.edu}

\author[0000-0003-3904-7378]{Logan Pearce}
\affiliation{Department of Astronomy, 
University of Michigan, 
Ann Arbor, MI 48109, USA}
\email{elapearce@umich.edu}

\author[0000-0003-4714-1364]{James G.~Ingalls}
\affiliation{IPAC, California Institute of Technology,
1200 E. California Boulevard,
Pasadena, CA 91125, USA}
\email{ingalls@ipac.caltech.edu}

\author[0000-0001-8014-0270]{Patrick Lowrance}
\affiliation{IPAC, California Institute of Technology,
1200 E. California Boulevard,
Pasadena, CA 91125, USA}
\email{lowrance@ipac.caltech.edu}

\begin{abstract}

Rings are ubiquitous among the giant planets in the Solar System, yet their presence around exoplanets remains observationally unconstrained. 
As direct imaging missions such as the \textit{Nancy Grace Roman Space Telescope} approach first light, understanding the influence of circumplanetary rings on reflected light photometry is critical for the accurate interpretation and follow-up of direct imaging observations. 
We present a hybrid forward model for reflected light phase curves of ringed exoplanets, coupling a ray casting framework for computing mutual planet-ring shadowing with empirical reflectance spectra from Cassini observations of Saturn's rings. 
We model Jupiter-like planets with opaque, icy rings spanning $1 - 10\times$ the radial extent of Saturn's A, B, and C rings across a grid of inclinations and ring obliquities. 
Our simulations show that rings enhance the flux ratios at favorable ring geometries. 
Saturn-equivalent rings increase peak flux by a factor of $3.1$ in the visible, while more extended rings produce enhanced contrasts exceeding an order of magnitude more.
The flux contribution of a ring is highly sensitive to obliquity and orbital location. 
At certain geometries, rings can suppress total system brightness below that of a ringless planet. 
We highlight multi-band observations as a complementary route for diagnosing the presence of an exoring. 
We also apply the model to $\upsilon$ Andromedae d as an observational test case and discuss the implications for exoring detectability with the Roman Coronagraph.
Exorings have the potential to bias inferred planetary properties from direct images and must be properly accounted for in observation preparation and interpretation.

\end{abstract}


\section{Introduction} \label{sec:intro}

    In recent years, exoplanet observations
    have revealed a great diversity of planetary system phenomena, including a potential ring system in a transit light curve around J1407b \citep{Kenworthy2014-vi, Kenworthy2015-xd}, a detected circumplanetary disk structure around PDS 70c \citep{Isella2019-xd, benisty2021ApJ...916L...2B}, and observations of low-density exoplanets which could be explained by the presence of ring systems (e.g., \citealt{piro2020AJ....159..131P}). 
    In the near future, direct imaging missions will continue to paint a more comprehensive picture of the exoplanet population as we reach toward imaging cooler, more mature planets in reflected light, and may begin to build a census of exoplanetary characteristics for Solar System analogs, including planets with circumplanetary ring systems. 
    
    The \textit{Nancy Grace Roman Space Telescope} (Roman) \blue{launched on} August 30, 2026 with a Coronagraph Instrument (hereinafter, CGI) technology demonstration that will perform high-contrast direct imaging of exoplanets from space \citep{Mennesson2020-jm, Poberezhskiy2022-dp}.
    The Coronagraph Instrument on Roman will enable direct detection of objects at separations of 3 -- 9 $\lambda$/D with flux ratios of 10$^{-8}$ or better, and may provide us with the first direct images of planets in reflected light \citep{bailey2023nancy}. 
    
    Previous studies have examined the nature of exoplanets in reflected light direct imaging in various contexts. 
    % While we do not provide an exhaustive review of the field of reflected light planet modeling here, we highlight a couple Roman-oriented results.
    For example, \cite{Batalha2018-gk} modeled giant planets in reflected light to understand how multi-band photometry may support first-order characterization with CGI (formerly \textit{WFIRST}-CGI). 
    They found that cloud coverage will significantly influence interpretation of the observations. 
    \cite{Lacy2019-kt} and \cite{Lacy2020-nv} used the \textit{CoolTLusty} model to further investigate the science that could be done with CGI. 
    Their models suggest that we may be able to roughly constrain the metallicity and atmospheric constituents of giant planets in reflected light. 
    Investigating self-luminous planets at optical wavelengths with CGI over multiple observation epochs will also support breaking the albedo-radius degeneracy inherent to direct imaging observations \citep{Lacy2020-nv}. 
    \blue{Additional work, such as \cite{Lupu2016AJ....152..217L} and \cite{Nayak2017PASP..129c4401N} investigated the ability to retrieve planetary properties (e.g., atmospheric abundances, gravity, clouds) and found that low signal-to-noise ratio, inadequate cloud models, unknown planet radius, and poorly-known orbital phase angles reduce the accuracy of the retrieved parameters.}
    However, these models also assumed giant planets with no circumplanetary material, which could significantly alter the planets' blended spectral energy distribution and further bias the interpretation of such observations.
    
    Reflected light photometry provides a promising opportunity to directly detect and characterize rings around an exoplanet (hereinafter ``exorings''). 
    Light reflected by the unresolved exoplanet and ring system will vary with orbital phase and ring orientation \citep{Arnold2004-gc,Dyudina2005-zy}. 
    Rings have the potential to introduce distinctive photometric signatures to the phase curve, and may vary considerably from the phase curve of a ringless planet \citep{Arnold2004-gc,Dyudina2005-zy}. 
    Additional modeling efforts have shown that polarimetry may also aid in identifying and characterizing exorings \citep{Lietzow2023A&A...671A.113L, Veenstra2025A&A...693A.310V}. 
    
    In the Solar System, Saturn's rings outshine the planet in methane absorption bands \citep{Miller2024SSRv..220...70M}. 
    Using Cassini observations of Saturn as an analog, \cite{Coulter2022ApJS..263...15C} showed that rings inclined by $20^\circ$ for an observation near quadrature (phase angle $\sim90^\circ$) would introduce a 30$\%$ increase in reflectivity in the planetary spectrum over visible wavelengths.
    Photometric variation due to exorings may be detectable in multi-epoch imaging that samples the phase curve of a ringed giant planet in reflected light with future observatories.
    % However, to date, no studies have explored the detectability of exoplanet rings with observatories like Roman or the Habitable Worlds Observatory in detail. 
    % \blue{cite \citep{Bowens-Rubin2025AJ....170..284B}}

    To date, detections of exorings remain elusive likely due to observational biases and the difficulty of detection. 
    However, rings are present around all of the giant planets in the Solar System and it is therefore reasonable to assume that rings will persist around exoplanets.
    We are motivated by three main points: the potential unknowns that may be uncovered by CGI, the physical phenomena that we may find when searching for exoplanets, and the inherent challenge to interpreting photometric detections that can depend heavily on modeling assumptions. 
    We investigate the potential for identifying the presence of exorings in multi-epoch imaging with future high-contrast imaging observatories like CGI \citep{Mennesson2020-jm}, the \textit{Habitable Worlds Observatory} (HWO, \citealt{Feinberg2024SPIE13092E..1NF}), the \textit{Giant Magellan Telescope}  \citep{Males2024SPIE13096E..0YM}, and the \textit{Extremely Large Telescope} (ELT; \citealt{Kasper2021Msngr.182...38K}). 

    We aim to answer the following three questions: 
    Could we detect an opaque, icy ring of some size around an exoplanet?
    Under what geometries do opaque rings most significantly alter an exoplanet's photometry?
    How do exorings bias the interpretation of direct images of planetary systems?
    We develop a ringed exoplanet simulator to model a planet's fraction of illumination visible to an observer due to mutual planet-ring shadowing and arbitrary orbital geometry. 
    We apply the model to a Jupiter-like planet with opaque, icy rings to model exorings as a function of size, orbital inclination, and ring obliquity. 
    We then explore a test case system to be observed by Roman, $\upsilon$ Andromedae d, to understand the impact on multi-epoch observations that would be brought by exorings.

    The rest of this paper is organized as follows. 
    In Section~\ref{sec:methods}, we describe the ringed planet model and detail the chosen parameters and spectra used to generate the model phase curves. 
    We present the model phase curves as a function of size, obliquity, and color in Section~\ref{sec:phase_curves}. 
    In Section~\ref{sec:upsAnd_caseStudy}, we apply the model to a case study system that will be observable with CGI, $\upsilon$ Andromedae d. 
    Section~\ref{sec:discussion} details our discussion of the results and implications for detection with Roman and future high-contrast imaging observatories.
    We summarize our key findings in Section~\ref{sec:summary}. 

\section{Methods} \label{sec:methods}

\subsection{Ringed Planet Model}

\begin{figure}
    \centering
    \includegraphics[width=0.98\linewidth]{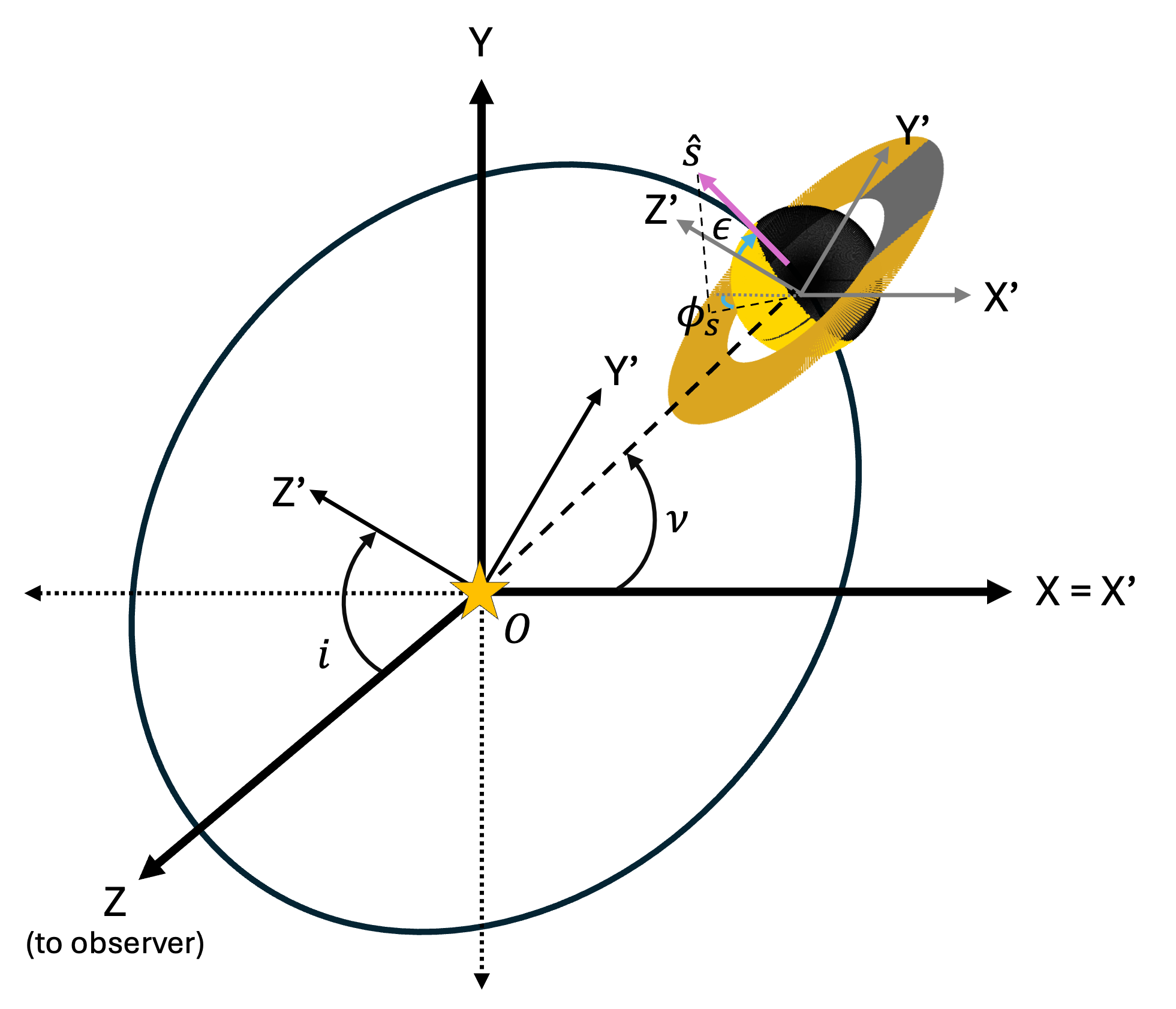}
    \caption{Diagram of the coordinate system used in the ringed planet model. The host star is located at the origin of the coordinate system, with the observer along the $Z$-direction and the $(X, Y)$ plane corresponding to the plane of the sky. The planet's ring  is tilted relative to the orbital plane with an obliquity, $\epsilon$, given by the angle between $Z'$ and the planet's spin axis, $\hat{s}$.
    The true anomaly, $\nu$, describes the location of the planet along its orbit.}
    \label{fig:ref_frames}
\end{figure}

We simulate the spectrum of a ringed gas giant as a function of orbital phase by building a hybrid model that couples a ray casting framework with empirical reflectance spectra. 
We adopt a Sun-like host star as a point source located at the origin of the coordinate system with a Solar spectral energy distribution\footnote{Reference spectrum obtained from STScI, \url{ftp://ftp.stsci.edu/cdbs/calspec}, 2024 December.}. 
The observer is placed along the $Z$-axis at a distance of 10\,parsecs. 
The planet and its ring are positioned at each epoch according to its Keplerian orbital elements and ring parameters.
The output of the geometric component of the model is the illuminated portion of the planet and ring that is visible to the observer at each observed orbital phase. 
The fractional illuminated areas are later used as multiplicative scaling factors applied to the corresponding empirical reflectance spectra to produce the total system planet-to-star flux ratio ($F_\mathrm{total}/F_\mathrm{\star}$). 
The parameters used in the ringed planet model are given in Table~\ref{tab:model_parameters}.
A diagram of the model coordinate system is shown in Figure~\ref{fig:ref_frames}.

\begin{table}[b]
    \centering
    \caption{Parameters and their definitions for the ringed exoplanet model.}
    \label{tab:model_parameters}
    \medskip
    \begin{tabular}{cl}
    \hline
    \textbf{Parameter} & \textbf{Definition} \\
    \hline
    \multicolumn{2}{l}{Orbit Parameters:} \\
        $a$ & Semi-major axis \\
        $e$ & Eccentricity    \\
        $i$ & Inclination     \\
        $\omega$ & Argument of periapsis \\
        $\Omega$ & Longitude of ascending node \\
        $\nu$ & True anomaly \\ 
        $\alpha$ & Phase angle \\
        
    \multicolumn{2}{l}{Spectral Parameters:} \\
        $A_g(\lambda)$ & Geometric albedo spectrum \\
        $\Phi(\alpha)$ & Phase function \\
        
    \multicolumn{2}{l}{Planet Parameters:} \\
        $R_p$ & Planet radius \\
        $\hat{s}$ & Spin axis \\
        $f_{\rm planet}(\alpha)$ & Illuminated planet fraction \\
        
    \multicolumn{2}{l}{Ring Parameters:} \\
        $\epsilon$ & Obliquity \\
        $\phi_s$ & Spin longitude \\
        $R_{\rm in}$ & Ring inner radius \\
        $R_{\rm out}$ & Ring outer radius \\
        $f_{\rm ring}(\alpha)$ & Illuminated ring fraction \\
        
    \hline
    \end{tabular}
\end{table}

\subsubsection{Coordinate System and Geometry}\label{sec:coordsystem}

The ringed planet model is defined in two reference frames (see Figure~\ref{fig:ref_frames}): the observer's reference frame in the plane of the sky, denoted by $(X, Y, Z)$, and the orbital frame, denoted by $(X', Y', Z')$. 
The host star is located at the origin \blue{of the observer's reference frame} and the observer is located along the $Z$-axis. 
The orbit is described by the Keplerian orbital elements: the semi-major axis ($a$), eccentricity ($e$), inclination ($i$), longitude of ascending node ($\Omega$), argument of periastron ($\omega$), and true anomaly ($\nu$). 
The inclination is the standard rotation of the orbital plane about the $X\mathrm{=}X'$-axis measured between the orbit normal and the sky-plane normal. 
$\nu$ represents the location of the planet along its orbit and serves as a proxy for orbital phase in this work. 
In the model reference frame, $\nu=0^\circ$ is defined when the planet is along the positive $X$-direction and reflecting at quadrature, $\nu=90^\circ$ is maximum illumination (superior conjunction, if $i=90^\circ$), $\nu$ \blue{between $0$--$90^\circ$ and} $90$--$180^\circ$ are gibbous phases, and $\nu=270^\circ$ is equivalent to minimum illumination (inferior conjunction, if $i=90^\circ$). 
The phase angle, $\alpha$, at each epoch is defined as the angle subtended at the planet between the directions to the star ($\hat{\mathbf{r}}_\star$) and to the observer ($\hat{\mathbf{Z}}$):

\begin{equation}
    \cos\alpha = \hat{\mathbf{r}}_{\star} \cdot \hat{\mathbf{Z}}.
\end{equation}

The orientation of the planet is described by its spin axis, $\hat{s}$. 
The spin axis defines the direction of the planet's pole and is assumed to remain fixed in inertial space throughout the orbit. 
In the model presented here, the ring normal is equivalent to the spin axis.

The orientation of the ring is described by the obliquity, $\epsilon$, and the spin longitude, $\phi_s$. 
We define $\epsilon$ as the angle between the spin axis, $\hat{s}$ (equivalent to the ring normal), and the orbital angular momentum vector (equivalent to $\hat{Z}'$). 
$\epsilon$ specifies the tilt of the ring relative to the orbital plane. 
$\phi_s$ describes the azimuthal direction of the spin axis tilt within the orbital reference plane and is the angle between the projection of the ring normal onto the orbit plane and the $Y$-direction. 
The primary orientation angles contributing to the flux of the ringed planet in the model are the inclination, true anomaly, ring obliquity, and spin longitude.
The radial extent of the ring is defined by its inner and outer radii, $R_{\rm in}$ and $R_{\rm out}$, respectively. 
The ring lies in a plane perpendicular to the ring normal and centered on the planet. 
The ring normal is held fixed in inertial space throughout the orbit; precession of the ring plane is not yet considered in our model.

\subsubsection{Flux Ratio Calculation}

We model the phase-angle dependence of the planet's reflected flux with a Lambertian phase function \citep{lambert1760photometria, cahoy2010exoplanet}.
For an isolated, unobscured planet, the planet-to-star flux ratio is

\begin{equation}\label{eqn:FpFs_standard}
    \frac{F_p}{F_\star} = A_g(\alpha, \lambda)\;\Phi(\alpha)\left(\frac{R_p}{r}\right)^2,
\end{equation}

\noindent where $r$ is the planet-star separation, $A_g(\alpha,\lambda)$ is the geometric albedo, $\Phi(\alpha)$ is the phase function, $\alpha$ is the phase angle, and $\lambda$ is the wavelength.

When a ring system is present, the ring may block or cast a shadow over a portion of the planetary disk, reducing its apparent brightness. 
We capture this shadowing effect with $f_{\rm planet}(\alpha)$, a parameter that describes the illuminated fraction of the visible planetary disk.
The planet-star flux ratio then becomes

\begin{equation}\label{eqn:FpFs}
   \frac{F_p}{F_\star} = A_{g,p}(\alpha, \lambda)\;\Phi(\alpha)\;f_{\rm planet}(\alpha)\left(\frac{R_p}{r}\right)^2.
\end{equation}

In the case of a ringless planet, $f_{\rm planet}(\alpha)=1$, and Equation~\ref{eqn:FpFs} is equivalent to Equation~\ref{eqn:FpFs_standard}. 

The ring contributes an additional reflected component to the total system brightness. 
The ring's contribution ($F_r$) to the total flux ratio of the planet-ring system is

\begin{equation}\label{eqn:FrFs}
    \frac{F_r}{F_\star} = A_{g, r}(\alpha, \lambda)\;f_{\rm ring}(\alpha)\;\frac{(R_{\rm out}^2 - R_{\rm in}^2)\cos\epsilon}{r^2},
\end{equation}

\noindent where $\epsilon$ is the obliquity and $R_{\rm in}$ and $R_{\rm out}$ are the ring inner and outer radii, respectively. 
The factor $f_{\rm ring}(\alpha)$ accounts for the fraction of the ring in the planet's shadow, the orientation of the ring face with respect to the observer, and the occultation of ring points by the planetary disk (see Section~\ref{sec:raycasting}). 
$f_{\rm ring}(\alpha)$ stands in as a ring phase function in this work, since we do not yet consider additional ring scattering parameters.
In this work we consider only fully opaque rings for simplicity and leave ring transmissivity for future model improvements, therefore only direct reflection from the star-illuminated face contributes to the reflected flux.

The total reflected-light flux ratio at each orbital epoch is obtained by combining the planet and ring contributions:

\begin{equation}\label{eqn:F_total}
    \frac{F_{\rm total}}{F_\star} = \frac{F_p}{F_\star} + \frac{F_r}{F_\star}.
\end{equation}

Both terms are evaluated at the same instantaneous planet-star separation and phase angle. 
We sample equally spaced true anomaly (i.e., phase) values over a full orbit to generate each set of phase curves, spanning the full range of observable phase angles.

To determine the observable flux ratio, we incorporate realistic filter curves (see Section~\ref{sec:roman_filters}). 
We convert the total modeled flux ratio spectrum (Equation~\ref{eqn:F_total}) to a band-averaged flux ($F_{\rm total, B}$) by integrating over the filter transmission curve, $T(\lambda)$, with:

\begin{equation}
    F_{\rm total, B} = \frac{\int (F_\mathrm{total}/F_\star)\, T(\lambda)\, d\lambda}{\int T(\lambda)\, d\lambda}.
\end{equation}

\subsubsection{Determining Planet and Ring Illumination with Ray Casting}\label{sec:raycasting}

To determine which portions of the planet and ring are illuminated by the star and visible to the observer at a given epoch, we separate the planet and ring surfaces into ensembles of points over which ray intersection tests are performed. 
The planet is represented as a sphere of radius $R_p$, sampled at $N_\theta \times N_\phi$ uniformly spaced points (where $\theta$ and $\phi$ are the polar and azimuthal angles, respectively). 
The ring is an annulus sampled at $N_r \times N_\phi$ points distributed uniformly in radius between the inner and outer ring radii ($R_{\rm in}$ and $R_{\rm out}$), where $N_r$ is the number of radial divisions and $N_\phi$ is the number of azimuthal divisions in the ring.
For the planets modeled here, we use a sphere of $N_\theta=260$ by $N_\phi=260$ points for the planet and an annulus of $N_r=260$ by $N_\phi=360$ for the ring.
These values were chosen through trial-and-error testing to balance computational runtime with sufficient spatial resolution to produce adequate sampling of the planet and ring. 

To determine whether a point on the planet, $\mathbf{p_\mathrm{planet}}$, is illuminated with respect to the observer, we evaluate three conditions. 
First, the point must lie on the observer-facing hemisphere. 
Second, the ray cast from the host star toward the planet must not intersect the planetary sphere at a closer distance than the point $\mathbf{p}$ (this implies self-shadowing, which removes points on the planet's night side). 
Third, the same ray must not intersect the ring annulus at a closer distance (this implies a ring shadow). 
A planet point is considered to be illuminated and observable only if it satisfies all three conditions. 
The illuminated fraction of the visible planet disk, $f_{\rm planet}(\alpha)$, is the fraction of observer-facing points that meets these conditions.

\begin{figure}[h!]
        \centering
        \includegraphics[width=0.8\linewidth]{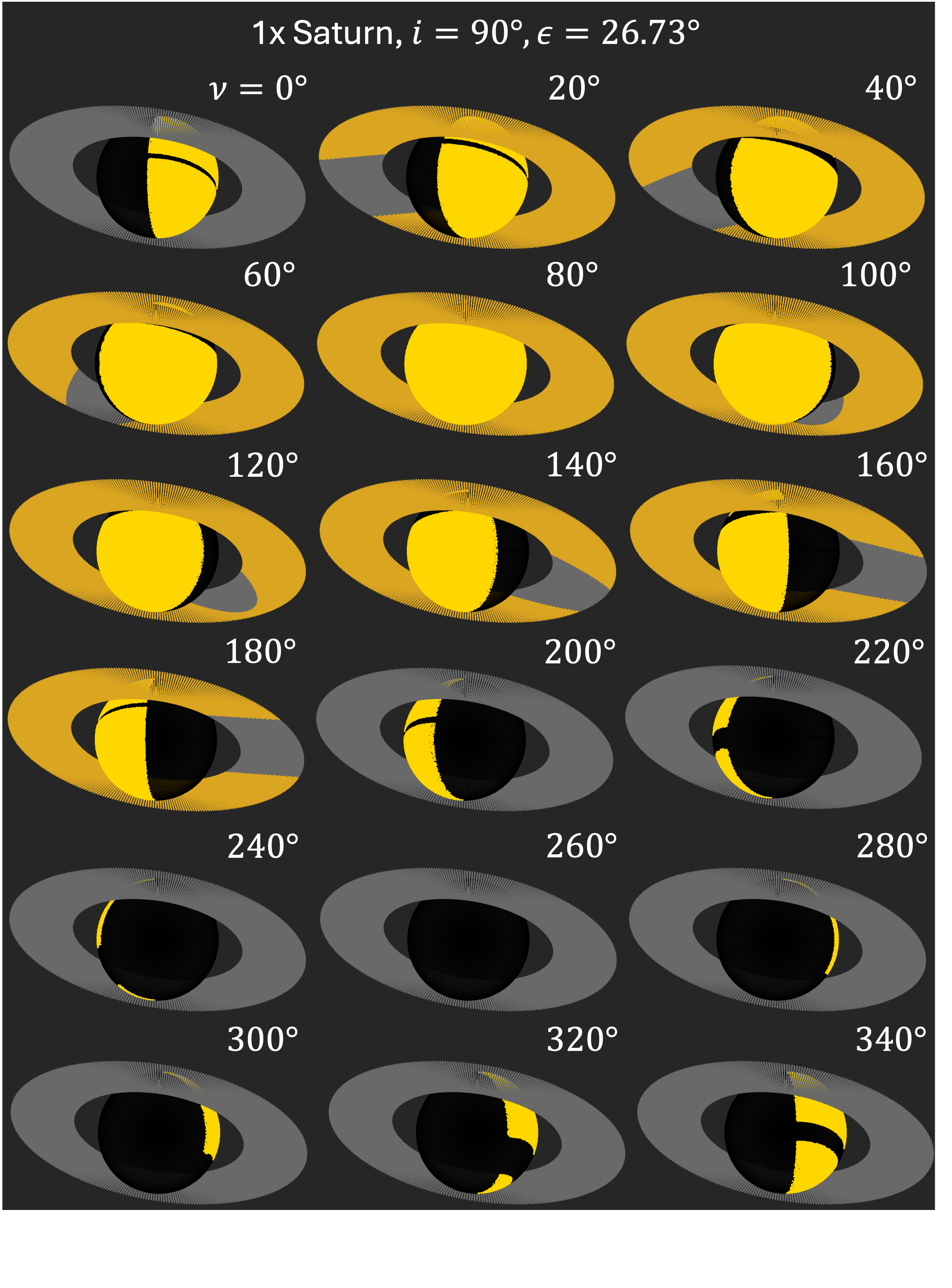}
        \caption{Output of the ringed planet model for different orbital positions ($\nu$) for the base case model, with $i=90^\circ$ and $\epsilon=26.73^\circ$. Points illuminated with respect to the observer are shown in gold and shadowed points are shown in gray/black. During the second half of the orbit ($\nu>180^\circ$), the side of the ring facing the observer is no longer illuminated. The corresponding phase curve is given in Figure~\ref{fig:ringed_vs_ringless_1-10x}.}
        \label{fig:planetModel_1xSat}
    \end{figure}

We determine the ring illumination and shadowing with an analogous set of tests for each ring point, $\mathbf{p_\mathrm{ring}}$.
First, the ray from the star must not be intercepted by the planetary sphere before reaching the ring (i.e., the planet's shadow is cast onto the ring). 
Second, the ring surface normal and the observer direction must lie on the same side of the ring plane, ensuring the observer views the illuminated face of the ring. 
Third, the line-of-sight ray from $\mathbf{p_\mathrm{ring}}$ toward the observer must not be occulted by the planet. 
The observable illuminated ring fraction, $f_{\rm ring}(\alpha)$, is the fraction of ring sample points satisfying these three criteria. 
The output of the ray-casting framework is $f_{\rm planet}(\alpha)$ and $f_{\rm ring}(\alpha)$, which are used as the multiplicative scaling factors in Equations~\ref{eqn:FpFs} and~\ref{eqn:FrFs}.
\blue{A visualization of the output is presented in Figure~\ref{fig:planetModel_1xSat} for a test case system.}

\subsubsection{Planet and Ring Reflectance Spectra}\label{sec:spectra}

For our giant planet spectrum, we adopt the empirically measured, disk-integrated albedo spectrum of Jupiter from \cite{Karkoschka1998Icar..133..134K}, which spans 300 to 1050\,nm and captures the visible-wavelength reflectance of the giant planet atmosphere and the characteristic methane absorption features. 
This spectrum serves as the geometric albedo, $A_{g,p}$, in Equation~\ref{eqn:FpFs}. 
We apply the Lambertian phase function and illuminated planet fraction derived from the ray-casting model to the geometric albedo spectrum to produce the phase- and wavelength-dependent flux ratio at each epoch. 

We assume an icy Saturn-like ring for our ringed planet model. 
The reflectance spectrum of the ring is taken from Cassini Visual and Infrared Mapping Spectrometer (VIMS) observations of Saturn's B ring from \cite{Hedman2013Icar..223..105H}.
The VIMS spectra cover two channels from 0.35 -- 1.05\,$\mu$m and 0.85 -- 5.20\,$\mu$m. 
For this work, we adopt the average spectra for wavelengths blueward of 1.0\,$\mu$m from observation RDHRCOMP001 for the B ring, as in \cite{Hedman2013Icar..223..105H}. 

\begin{figure}[h!]
    \centering
    \includegraphics[width=\linewidth]{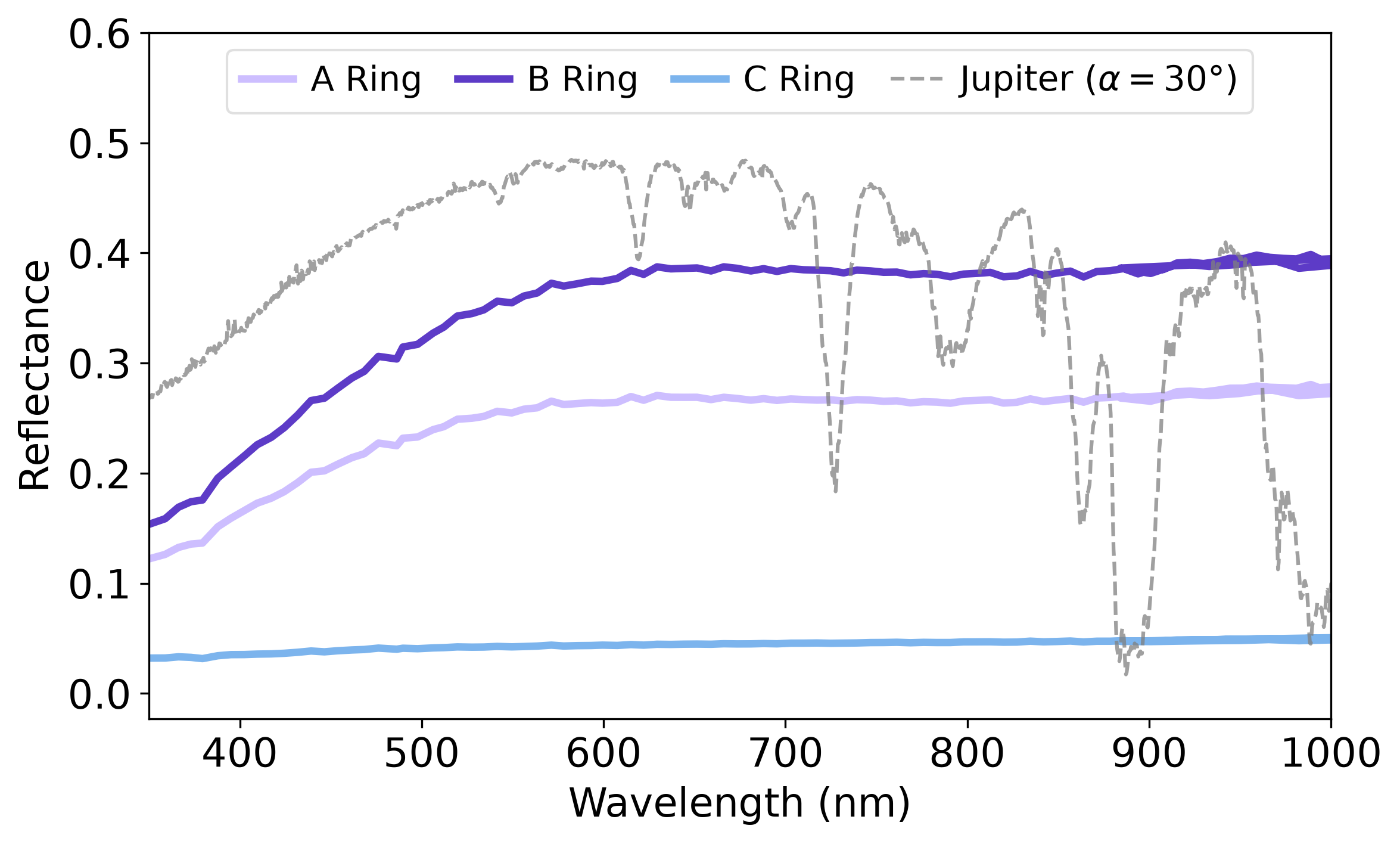}
    \caption{Reflectance spectra of Saturn's A, B, and C rings observed by Cassini/VIMS compared to the spectrum of Jupiter at $\alpha=30^\circ$ (dashed gray line). We adopt the B ring spectrum in our ring model as it has the highest optical depth of the three sections. The plotting artifacts at $\lambda \gtrsim900$\,nm are due to the overlapping wavelength coverage of the two VIMS channels.}
    \label{fig:ring_spectra}
\end{figure}

Saturn's ring spectra are dominated by water ice, creating highly reflective spectra at visible wavelengths (e.g., \citealt{Cuzzi2009sfch.book..459C, Hedman2013Icar..223..105H}).
We adopt a single spectrum shown in Fig.~\ref{fig:ring_spectra} for the full ring annulus.
Figure~\ref{fig:ring_spectra} also shows the individual average reflectance spectra of Saturn's A, B, and C rings.
\blue{Saturn's A, B, and C rings have extents ranging from $\sim1.3 - 1.6\,R_{\rm Saturn}$, $\sim1.6 - 2.03\,R_{\rm Saturn}$, and $\sim 2.1 - 2.4\,R_{\rm Saturn}$, respectively \citep{Colwell2009sfch.book..375C}.}
The B ring has the highest optical depth of the primary ring regions. 
\blue{The ring region that we adopt in this work (110,000 -- 111,000\,km) samples the central B ring, where the optical depth is the greatest ($3 \lesssim \tau \lesssim 6$; \citealt{Colwell2009sfch.book..375C, Cuzzi2018prs..book...51C}).}

\subsubsection{Planet and Ring Model Parameterization}\label{sec:ring_planet_parameterization}

Our base model planet is a Jupiter-like planet with a radius of $1\,R_\mathrm{Jup}$ and a Saturn-like opaque, icy ring. 
The ring's inner and outer radii are set to $1.43 \,R_p$ and $2.47\,R_p$, respectively. 
These boundaries approximately encompass the radial extent of Saturn's main ring system, from the inner C ring to the outer A ring, when expressed in units of planetary radii (e.g., \citealt{Colwell2009sfch.book..375C, Cuzzi2010Sci...327.1470C}). 
The inner radius also falls outside Jupiter's Roche limit. 
The rings are assumed to be composed primarily of water ice, consistent with the observed composition of Saturn's main rings \citep{Cuzzi2009sfch.book..459C} and are treated as fully opaque for the cases presented here ($\tau\gg 1$). 
Radial variations in the ring optical depth and ring gaps are not modeled; the annulus is treated as a uniform reflector characterized by the composite spectrum (see Section~\ref{sec:spectra}). 

To assess the sensitivity of reflected light photometry to ring extent and various geometries, we model the ring extent and orbital configurations summarized in Table~\ref{tab:model_cases}. 
The ``base'' case ($1\times$ Saturn) adopts ring boundaries that are equivalent in radius to Saturn's C through A rings, scaled to a Jupiter-radius planet ($1.43\,R_{\rm Jup}$ to $2.47\,R_{\rm Jup}$).
Three additional cases extend the outer radius of the ring to $2\times$, $5\times$, and $10\times$ Saturn's ring extent. 
These correspond to outer radii of $3.52\,R_{\rm Jup}$, $6.65\,R_{\rm Jup}$, and $11.87\,R_{\rm Jup}$, respectively.

Each configuration is evaluated over a grid of orbital inclinations and ring obliquities (Table~\ref{tab:model_cases}), sampling a range of viewing geometries. 
The spin longitude, $\phi_s$, was chosen to be Saturn-like \citep{Arnold2004-gc}.
For all model cases, the planet is placed at Jupiter's orbital separation of 5.2\,AU with zero eccentricity, well beyond the water-ice snow line of a Sun-like host-star \citep{hayashi1981PThPS..70...35H, Podolak2004M&PS...39.1859P}. 

The phase curves in this work are presented as a function of true anomaly rather than phase angle to cover the full orbit regardless of system inclination. 
We note that this convention is consistent with the ``orbital phase angle'' definition presented in the \cite{Arnold2004-gc} phase curves.

\subsection{Roman Coronagraph Filters}\label{sec:roman_filters}

\begin{figure}
        \centering
        \includegraphics[width=\linewidth]{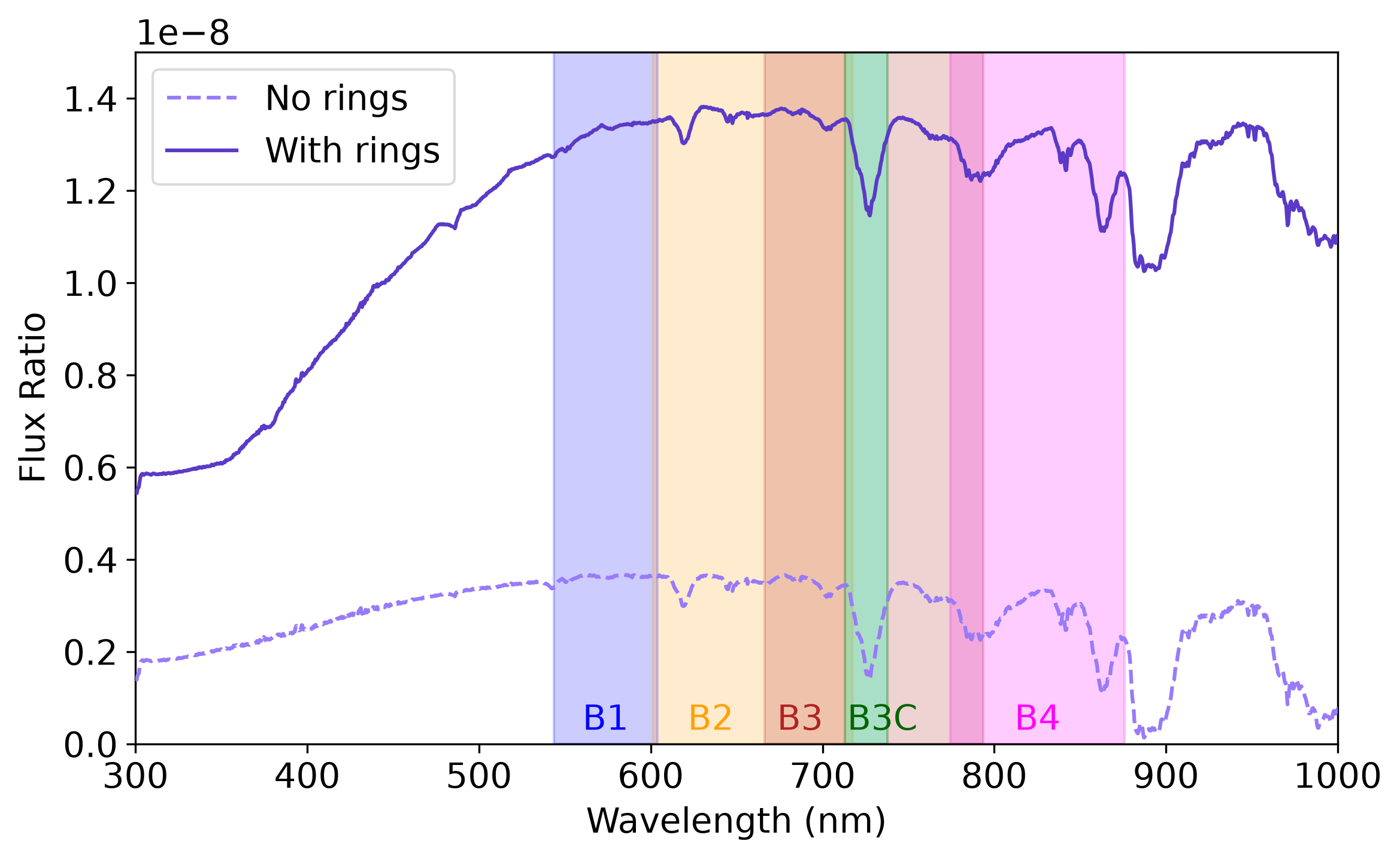}
        \caption{Spectra of Jupiter at full phase with (solid) and without (dashed) icy rings 1$\times$ the size of Saturn's. The planet spectrum is from \cite{Karkoschka1998Icar..133..134K}; the ring spectrum is Saturn's B ring spectrum from \cite{Hedman2013Icar..223..105H} (average measured phase angle of $\sim27^\circ$). Select CGI filters are overplotted as horizontal bands labeled B1, B2, B3, \blue{B3C,} and B4 for bandpasses 1, 2, 3, \blue{3C} and 4, respectively. CGI filter curves were measured by Materion and are available from the Roman Science Support Center at IPAC\footnote{Obtained from IPAC at Caltech, \url{https://roman.ipac.caltech.edu/page/additional-coronagraph-instrument-parameters-model-and-data-html\#Color_Filter_Curves}, 2024 December.}.}
        \label{fig:planet_rings_filters}
    \end{figure}

We adopt passbands from CGI to model the phase curves for this work \citep{bailey2023nancy}. 
CGI's fully-supported observing mode uses the Hybrid Lyot Coronagraph (HLC) imaging mode in Band 1 with a narrow field-of-view and central wavelength ($\lambda_c$) of $573.8$\,nm with a 9.8\% bandwidth ($\Delta \lambda$; see Figure~\ref{fig:planet_rings_filters}).
Band 1 spans a wavelength range in which the reflected light from a giant planet atmosphere and the high albedo of water-ice rings contribute substantially to the total system brightness, making it well-suited to probing for signatures of rings in the phase curves. 
We restrict this analysis primarily to Band 1, since it is the only fully supported observing mode of the coronagraph and covers visible wavelengths. 

We briefly touch on a multi-color analysis in Section~\ref{sec:phase_curves} using Bands 3 (729.3\,nm, $\Delta \lambda$=16.8\%), 4, and a narrow bandpass engineering filter, 3C (725.9\,nm, $\Delta \lambda=2.8\%$). 
Band 4 (825.5\,nm, $\Delta \lambda = 11.7\%$) is a filter that will be supported on a ``best effort'' basis, meaning it does not have guaranteed support on-orbit.
Additional filter and field-of-view combinations may become available as instrument performance requirements are met on orbit \citep{Riggs2021SPIE11823E..1YR, bailey2023nancy, Riggs2025JATIS..11b1403R}. %, and represent a natural extension of this work \blue{(see Section~\ref{sec:future_work})}. 

\begin{table*}
    \centering
    \caption{Summary of parameters for each of the ringed planet model cases.}
    \label{tab:model_cases}
    \begin{tabular}{cllll}
    \hline
    \textbf{Parameter} & \textbf{$1\times$ Saturn (Base)} & \textbf{$2\times$ Saturn} & \textbf{$5\times$ Saturn} & \textbf{$10\times$ Saturn} \\
    \hline
        $R_p$ & 1 $R_{\rm Jup}$ & 1 $R_{\rm Jup}$ & 1 $R_{\rm Jup}$ & 1 $R_{\rm Jup}$ \\
        $R_{\rm in}$ & 1.43 $R_p$ & 1.43 $R_p$ & 1.43 $R_p$ & 1.43 $R_p$ \\
        $R_{\rm out}$ & 2.47 $R_p$ & 3.52 $R_p$ & 6.65 $R_p$ & 11.87 $R_p$  \\
        
        $i$ & $30^\circ$, $60^\circ$, $90^\circ$  & $30^\circ$, $60^\circ$, $90^\circ$ & $30^\circ$, $60^\circ$, $90^\circ$ & $30^\circ$, $60^\circ$, $90^\circ$ \\
        
        $\epsilon$ & $26.73^\circ$, $45^\circ$, $90^\circ$ & $26.73^\circ$, $45^\circ$, $90^\circ$ & $26.73^\circ$, $45^\circ$, $90^\circ$ & $26.73^\circ$, $45^\circ$, $90^\circ$ \\
        
        $\phi_s$ & $-90^\circ$ & $-90^\circ$ & $-90^\circ$ & $-90^\circ$ \\
    \hline
    \end{tabular}
\end{table*}

    \begin{figure*}
        \centering
        \includegraphics[width=\linewidth]{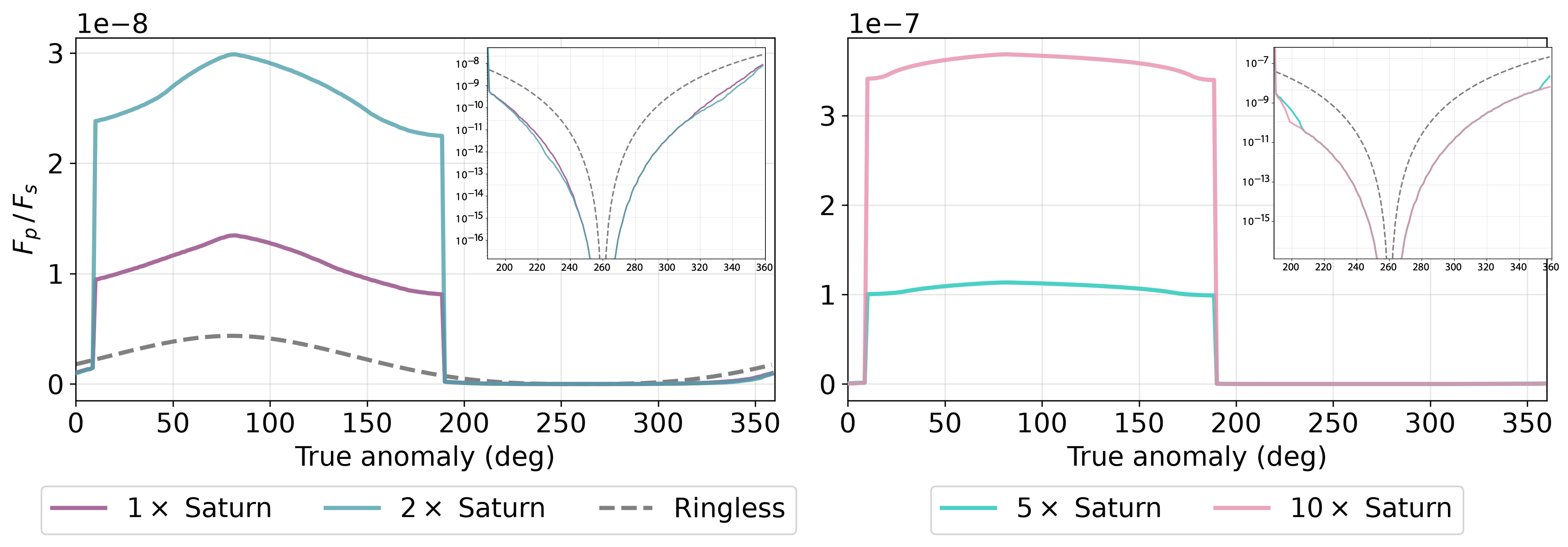}
        \caption{Phase curves of planets with rings (solid) of varying radial extent (1 -- 10$\times$ Saturn's rings, $\epsilon=26.73^\circ$, $i=90^\circ$) compared to a ringless planet (Lambertian; dashed line) in Roman CGI Band 1. Rings increase the total flux ratio of the system \blue{over the first half of the orbit} compared to a ringless planet, and alter the morphology of the phase curve through mutual shadowing. Because these models include an opaque ring that does not account for ring scattering properties, the ring is not illuminated with respect to the observer in the second half of the orbit ($\nu>180^\circ$) and casts a shadow onto the planet, reducing the total flux ratio. The plot insets show the minor differences in flux ratio in the second half of the orbit due to mutual shadowing by the various ring sizes.}
        \label{fig:ringed_vs_ringless_1-10x}
    \end{figure*}

\section{Phase Curves of Ringed Planets}\label{sec:phase_curves}

We present the modeled flux ratio for CGI's Band 1 as a function of true anomaly (that is, orbital location; see note on true anomaly versus phase angle in Section~\ref{sec:ring_planet_parameterization}) for four ring sizes alongside a ringless Lambertian baseline in Figure~\ref{fig:ringed_vs_ringless_1-10x}, evaluated at an orbital inclination of $90^\circ$ and ring obliquity of $26.73^\circ$ (Saturn-like; \citealt{Arnold2004-gc}). 
The presence of the ring system produces two characteristic effects on the reflected light phase curves in this opaque ring scenario: an enhancement of the total system brightness over the first half of the orbit, and an asymmetry in the shape of the phase curve. 

\begin{figure}
        \centering
        \includegraphics[width=0.8\linewidth]{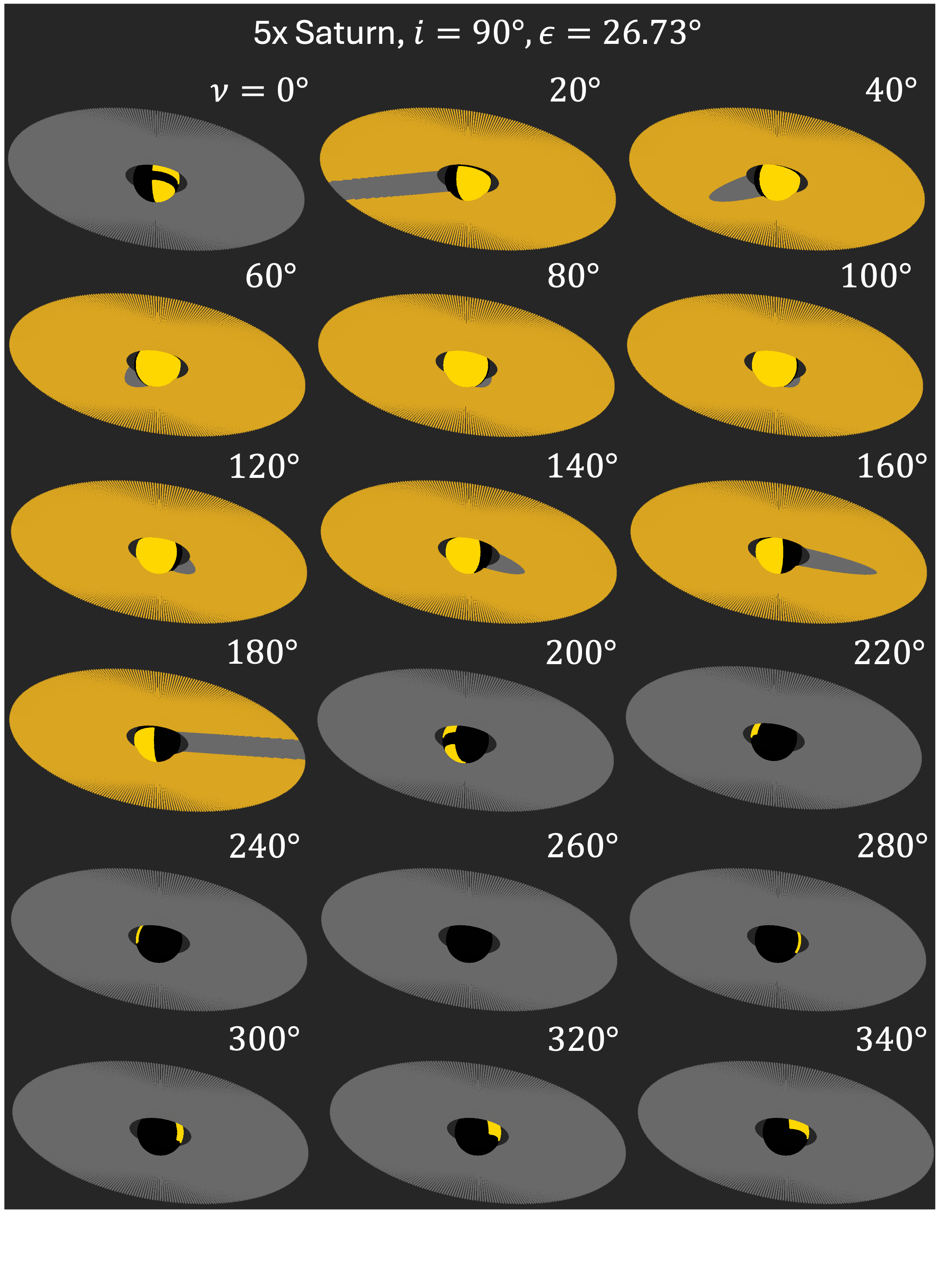}
        \caption{Same as Figure~\ref{fig:planetModel_1xSat} for the $5\times$ Saturn model. The corresponding phase curve is shown in Figure~\ref{fig:ringed_vs_ringless_1-10x}.}
        \label{fig:fig:planetModel_5xSat}
    \end{figure}

All four ring sizes have flux ratios brighter than the ringless planet during the first half of the orbit. 
The $1\times$ and $2\times$ Saturn configurations (Figure~\ref{fig:ringed_vs_ringless_1-10x}, left panel) reach peak flux ratios of $\sim1.3 \times 10^{-8}$ and $\sim2.99 \times 10^{-8}$ near $\nu = 90^\circ$, which corresponds to full illumination. 
The $5\times$ and $10\times$ Saturn configurations (Figure~\ref{fig:ringed_vs_ringless_1-10x}, right panel) reach peak flux ratios of $\sim1.1 \times 10^{-7}$ and $\sim3.7 \times 10^{-7}$ at full phase. 
This is in comparison to a peak flux ratio of $\sim4.4 \times 10^{-9}$ for the ringless planet. 
Model renderings in Figures~\ref{fig:planetModel_1xSat} and~\ref{fig:fig:planetModel_5xSat} illustrate the illumination of the planet and ring system along the orbit for the $1\times$ and $5\times$ Saturn cases.

\begin{figure}
        \centering
        \includegraphics[width=\linewidth]{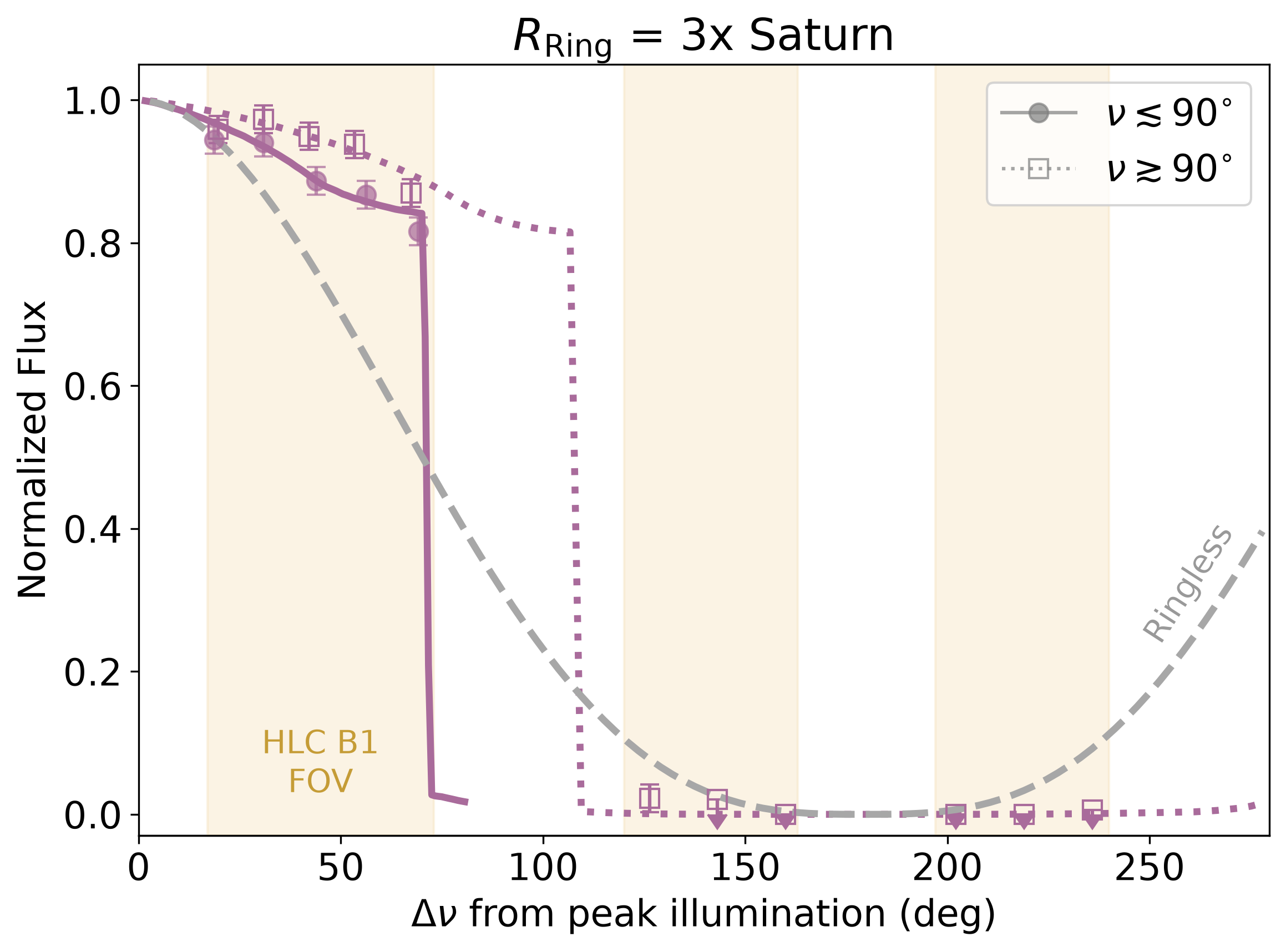}
        \caption{\blue{Band 1 phase-folded light curves for a ringed exoplanet with a ring 3$\times$\,the size of Saturn's (folded at the point of peak illumination; $i=90^\circ, \epsilon=26.73^\circ$). Phase-folding highlights the asymmetries in the phase curve due to mutual shadowing. The solid and dotted lines represent phase ($\nu$) values before and after peak illumination ($\nu=90^\circ$), respectively. Simulated observations where the planet would fall within CGI's HLC B1 field-of-view (FOV; gold) are also shown for $1{\rm e-}9$ photometric uncertainty. A ring $3\times$ the size of Saturn's would be large enough to distinguish the minor asymmetries in the phase curve with a photometric uncertainty of $1{\rm e-}9$.}}
        \label{fig:asymmetry}
    \end{figure}

The most conspicuous feature common to all configurations in the models is the steep drop in flux at the mid-point of the orbits.
The magnitude of the flux decrease and the flux level after the transition vary with ring size, with subtle variations shown in the inset panels of Fig.~\ref{fig:ringed_vs_ringless_1-10x}.
\blue{Some phase curves also exhibit smaller-scale asymmetries, as highlighted in the phase-folded curve in Figure~\ref{fig:asymmetry} for rings $3\times$ the size of Saturn's.}
We discuss these features in more detail in Section~\ref{sec:discussion}.

\begin{figure}
        \centering
        \includegraphics[width=\linewidth]{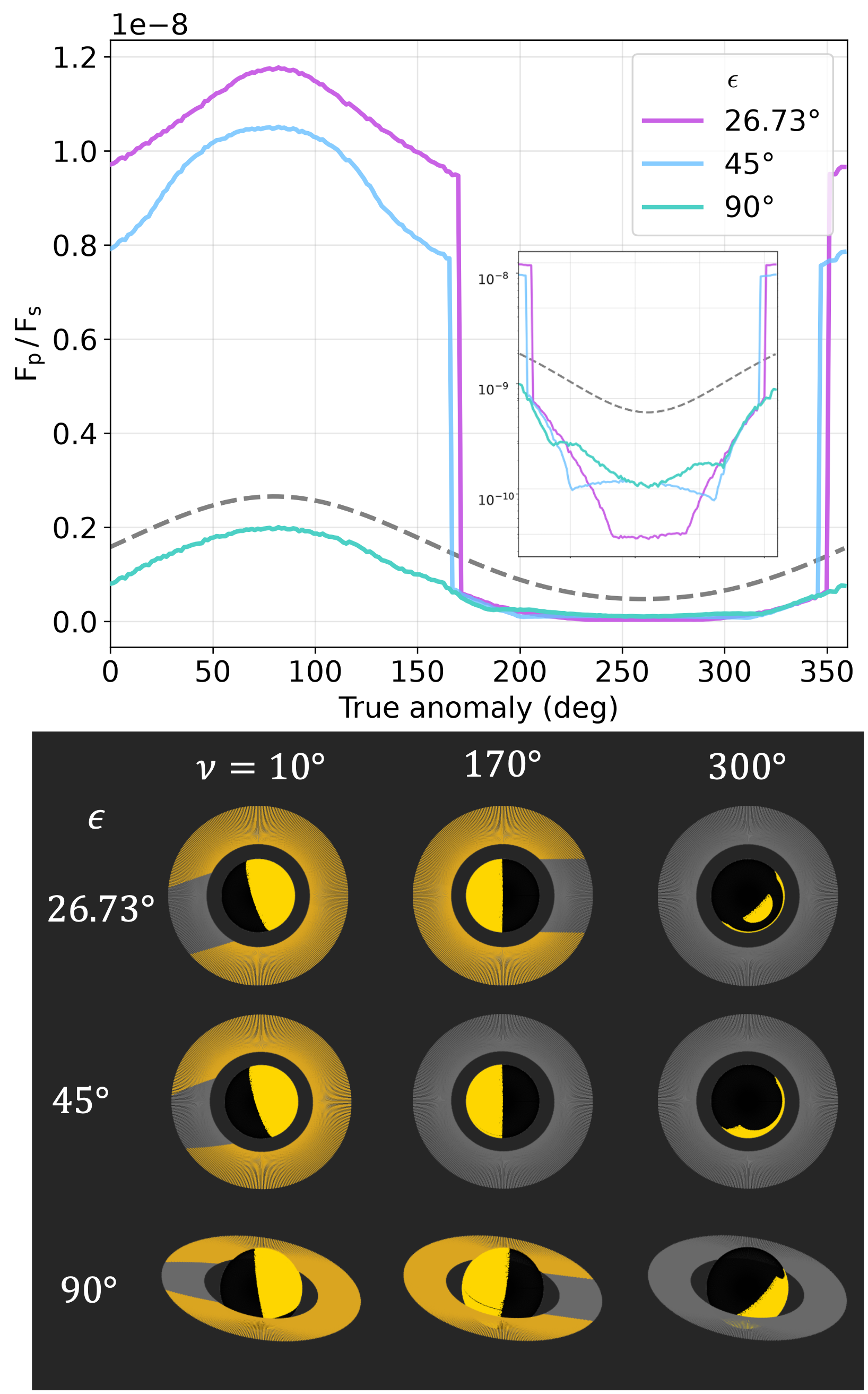}
        \caption{(Top) CGI Band 1 phase curves of $1\times$ Saturn ringed planets with $i=30^\circ$ for three obliquities compared to a ringless Lambertian planet (dashed line). 
        The obliquity of the ring introduces the distinct shape to the phase curve. At higher obliquities, the shadowing of the ring on the planet becomes more pronounced at certain geometries, and can decrease the overall system flux ratio to less than that of a Lambertian in the first half of the orbit due to the change in projected ring area. (Bottom) Same as Fig.~\ref{fig:planetModel_1xSat} and~\ref{fig:fig:planetModel_5xSat} for the phase curves shown above. Each row corresponds to $\epsilon=26.73^\circ, 45^\circ$, or $90^\circ$, respectively. Each column corresponds to $\nu=10^\circ, 170^\circ$, or $300^\circ$, respectively. }
        \label{fig:phasecurves_30incl_allObl}
    \end{figure}

We also examined the role of ring obliquity in influencing the shape and amplitude of the phase curve.
Figure~\ref{fig:phasecurves_30incl_allObl} shows phase curves for the $1\times$ Saturn configuration at $i=30^\circ$ for three obliquities: $26.73^\circ$, $45^\circ$, and $90^\circ$. 
As the obliquity ($\epsilon$) of the ring increases, the flux ratio of the planet is generally less compared to lower values of $\epsilon$ for the first half of the orbit.

\begin{figure}
        \centering
        \includegraphics[width=0.95\linewidth]{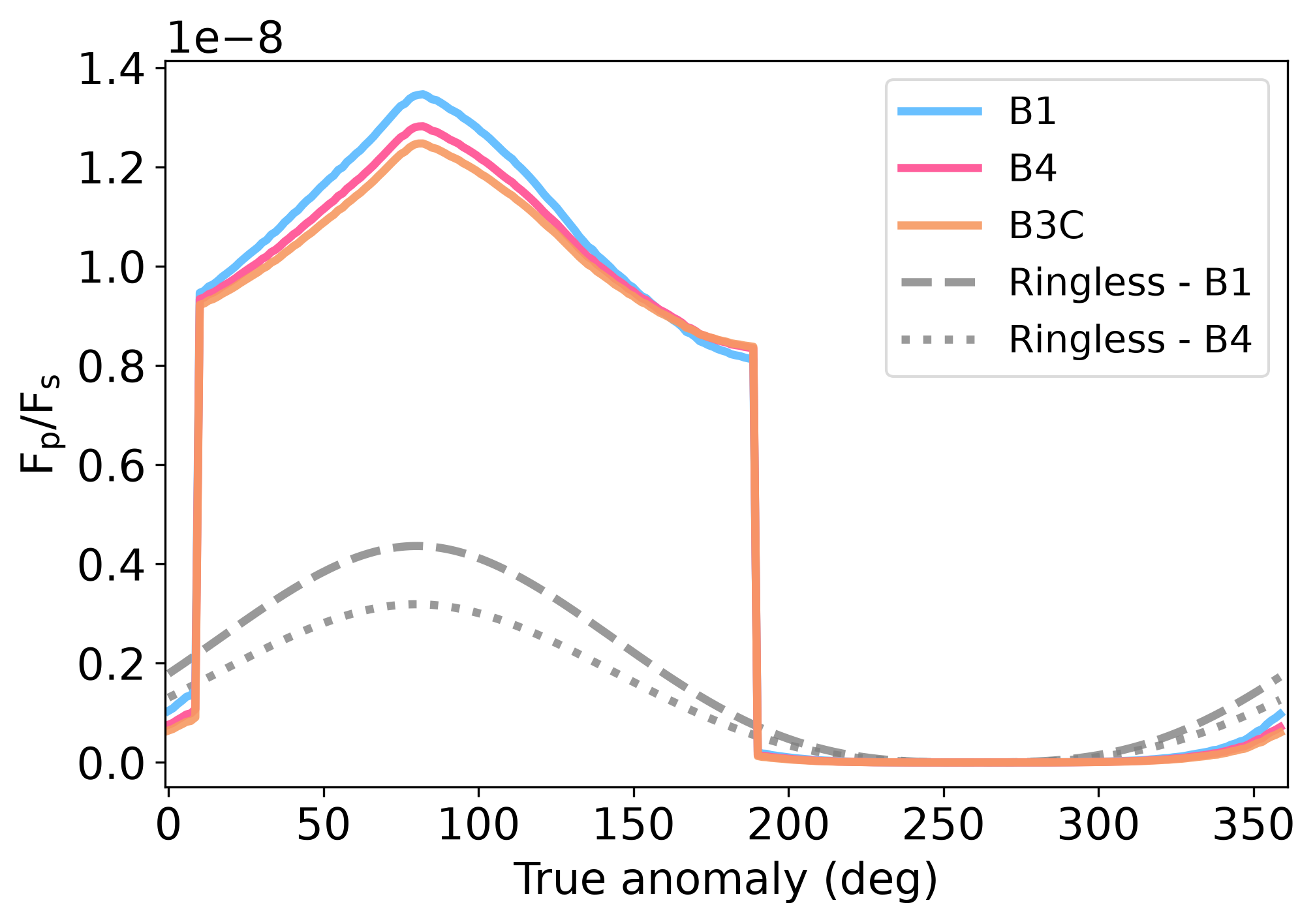}
        \caption{Phase curve color comparison in CGI Bands 1, 4, and an engineering narrow bandpass filter, 3C ($i=90^\circ, \epsilon=26.73^\circ, 1\times$ Saturn). Bands 4 and 3C cover regions of the planet's spectrum with deep methane features, which decrease the planet's total flux contribution. However, icy rings maintain their bright flux contribution in these regions.}
        \label{fig:phasecurves_color}
    \end{figure}

\begin{figure}
        \centering
        \includegraphics[width=0.75\linewidth]{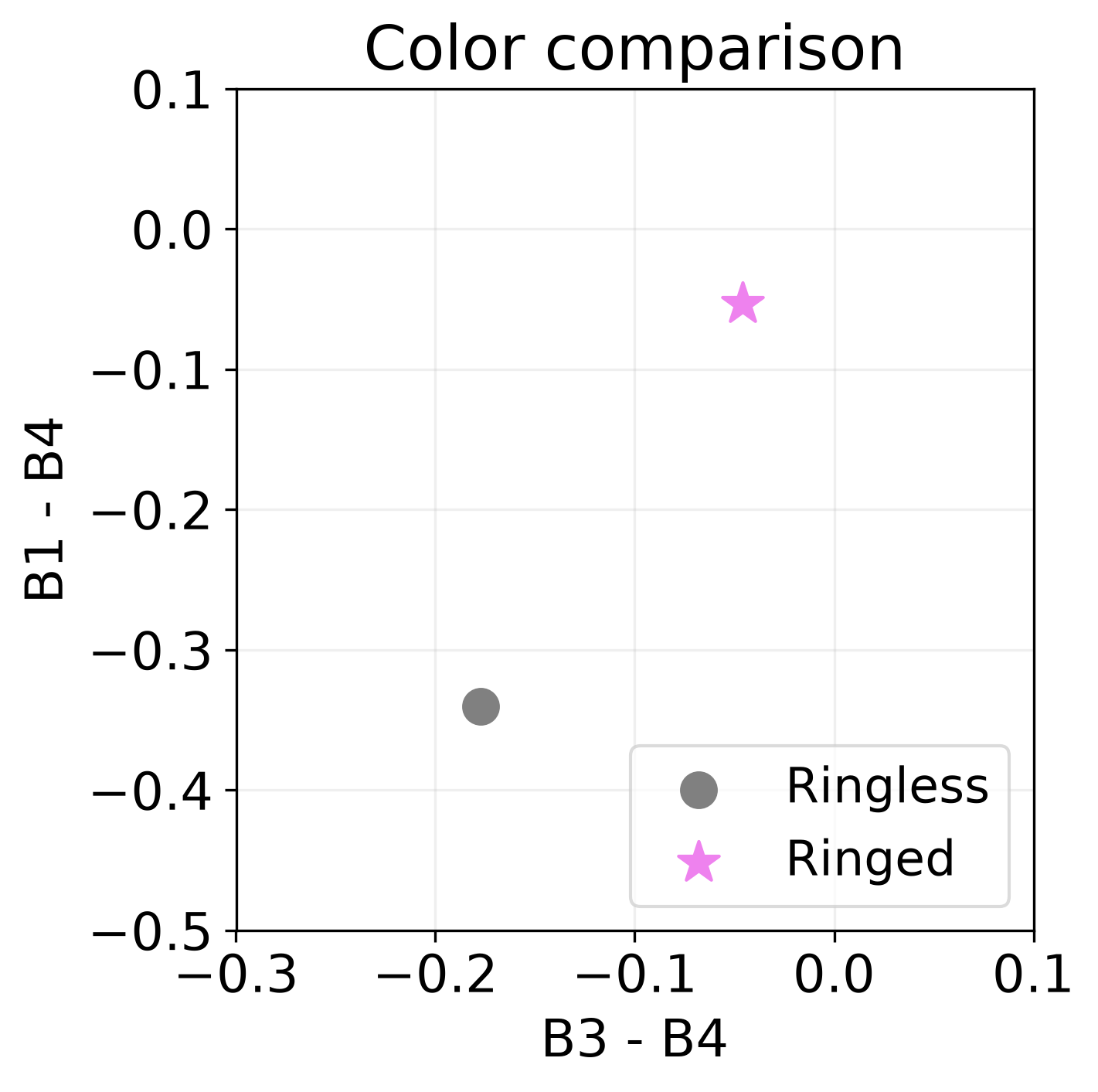}
        \caption{Color-color comparison of the $1\times$ Saturn ringed planet case presented in Figure~\ref{fig:phasecurves_color} compared to a ringless planet in CGI Bands 1, 3, and 4. The enhancement provided by the rings and absence of methane features in the ring spectrum pushes the ringed planet model to a distinct region of color space compared to the ringless planet.}
        \label{fig:color_comparison}
    \end{figure}

Multi-color observations offer a potential complementary method for diagnosing the presence of rings, as the spectra of giant planets and icy rings may contain different molecular features. 
Figure~\ref{fig:phasecurves_color} compares phase curves evaluated in Roman CGI Bands 1 and 4, and an engineering narrow bandpass filter, 3C.
The Bands 1, 4, and 3C phase curves reach peak flux ratios of $\sim1.35\times 10^{-8}$, $\sim1.28\times 10^{-8}$, and $\sim1.25\times 10^{-8}$, respectively. 
Figure~\ref{fig:color_comparison} shows the a color comparison between Bands 1, 3, and 4, where the ringed and ringless models occupy distinct regions of parameter space. 

\section{$\upsilon$ Andromedae d Test Case}\label{sec:upsAnd_caseStudy}

We also apply the ringed planet model to $\upsilon$ Andromedae d (hereafter $\upsilon$ And d) as a test case observational application, as it is a promising target for CGI \citep{Saxena2021AJ....162...30S}. 
$\upsilon$ Andromedae is an F9V star located 13.48\,pc from the Sun \citep{Abt2009ApJS..180..117A, Gaia2020yCat.1350....0G} with three confirmed planet companions \citep{Christiansen2025PSJ.....6..186C}. 
$\upsilon$ And d is the outer-most planet in the system. 
It is a giant planet with a semi-major axis of 2.513\,AU and an eccentricity of 0.2987 \citep{Curiel2011AA...525A..78C}, placing it in the range plausibly just beyond the water-ice snow line for a solar-type host. 
The orbital inclination, $i=23.76^\circ$, was measured via Hubble Space Telescope astrometry \citep{McArthur2010ApJ...715.1203M}, providing a relatively well-constrained orbital geometry from which the planet's projected separation and phase angle can be predicted as a function of time. 

We adopt a planetary radius of $1\,R_{\rm Jup}$ and ring extent of $1.43\,R_p$ to $2.47\,R_p$. Because the spin-axis orientation of $\upsilon$ And d is not observationally constrained, we test a three different obliquity values in our models.
All planet and model parameters are listed in Table~\ref{tab:upsAnd_parameters}.

$\upsilon$ And d is a well-suited test case for this study for several reasons. 
Its physical separation from the host star and proximity of the system produce projected angular separations ($\sim\,$120 -- 224\,mas; $\sim$2.4 -- 4.5\,$\lambda/D$) that fall within CGI's accessible field of view during portions of the mission timeline, making it one of the favorable known giant planet targets for Roman observations. 
The measured orbital inclination removes a key geometric degeneracy and allows us to make better-constrained predictions for the range of accessible phase angles than other known radial velocity targets with unconstrained inclination. 
We use this system to predict the planet-to-star flux ratio over a full orbital period with and without a ring, estimate favorable observing windows for Roman, and evaluate the degree to which a ring could bias the inferred planetary photometry.

We use the  ringed planet model to generate simulated phase curves for CGI's HLC Band 1 (see Section~\ref{sec:roman_filters}) of $\upsilon$ And d with and without a ring.
The ringless case adopts Jupiter's spectrum, scaled to the orbital distance of $\upsilon$ And d (we do not account for the difference in host star SED in this model).
We adopt the planet and ring parameters shown in Table~\ref{tab:upsAnd_parameters}, with a ring extent equivalent to that of Saturn's A, B, and C rings, consistent with the $1\times$ Saturn base case in Section~\ref{sec:phase_curves}.

We compute the flux ratio as a function of orbital location for a ringed and ringless $\upsilon$ And d over an orbital period and compare this to Roman's observing windows. 
Due to its on-sky location and orbital configuration, $\upsilon$ And d is observable immediately during the first six months of Roman's observing period (contingent on the launch date). %, due to its projected separation and the location on-sky. 
After early 2027, the planet falls within the inner working angle of the HLC and is not observable again until a year later. 

We used the \texttt{corgiETC} exposure time calculator \citep{corgietc} to determine the achievable signal-to-noise ratio (SNR) and corresponding integration times with the default setting for local zodiacal light ($6.3096\times10^{-10}$\,arcsec$^{-2}$) and assuming the ``optimistic'' instrument performance scenario for HLC Band 1. 
An SNR of 5 is achievable in 0.45\,hours and 3.1\,hours for flux ratio levels of $1\times10^{-8}$ and $5\times10^{-9}$, respectively. 
We compute the flux ratio uncertainty by dividing the flux ratio for an observation by the SNR and apply Gaussian noise to each observation. 

\begin{figure*}
    \centering
    \includegraphics[width=\linewidth]{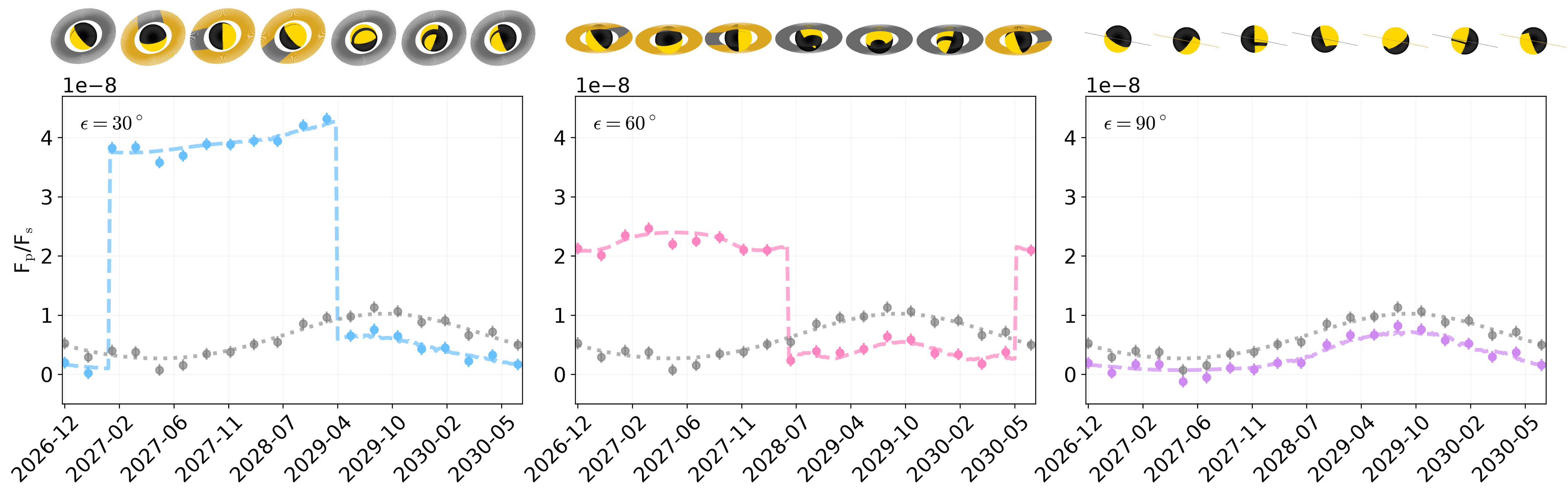}
    \caption{Simulated CGI Band 1 phase curves for $\upsilon$ And d with rings (color, dashed) and without (gray, dotted) for three different ring obliquities. Ringed planet models illustrating the geometry of the ring are shown above each respective plot for a few oribtal locations.
    With a ring, the maximum flux ratio achieved in the $\epsilon=30^\circ$ (left) and $60^\circ$ (center) cases is 4.2 and 2.3 times greater, respectively, pushing it above an SNR of 5 for flux ratio levels of $1\times10^{-8}$ and $5\times10^{-9}$. In the worst case scenario, a ring oriented edge-on to the observer ($\epsilon=90^\circ$; right) would contribute no additional flux while shadowing portions of the planet disk, thus decreasing the total system flux ratio below that of a ringless planet.}
    \label{fig:upsAnd_phasecurves}
\end{figure*}

The simulated observations with and without a ring for three different obliquities are shown in Figure~\ref{fig:upsAnd_phasecurves}. 
Without a ring, the planet only reaches a flux ratio of $\sim1\times10^{-8}$ near peak illumination. 
With a ring, the maximum flux ratio achieved depends on the ring obliquity, but can reach 4.2 times greater than the ringless case. 
This would push the system flux above the SNR of 5 threshold for both flux ratio levels, though the degree of enhancement is dependent on the ring obliquity. 
At $\epsilon = 30^\circ$, the ring boosts the peak flux to $\sim4.3\times10^{-8}$, well above the detection limit. 
At $\epsilon=60^\circ$, the contrast enhancement is $\sim2.4\times10^{-8}$, which is also sufficient to clear the SNR of 5 threshold from late 2026 through early 2028.  
At $\epsilon=90^\circ$, the ring is edge-on to the observer throughout the orbit and contributes no additional reflected flux.  
Instead, it shadows varying fractions of the planetary disk throughout the orbit, suppressing the total system flux ratio below that of the ringless case across the entire orbital period (Figure~\ref{fig:upsAnd_phasecurves}, right panel).
While this edge-on orientation would decrease the likelihood of detectability for Roman, HWO is expected to achieve flux ratios of better than $10^{-9}$ and would be capable of resolving the difference between the ringed and ringless cases given observations with low uncertainty. 

\begin{table}
    \centering
    \caption{Parameters for the $\upsilon$ And d and exoring phase curve model.}\label{tab:upsAnd_parameters}
    \footnotesize
    \begin{tabular}{ccl}
    \hline
    \textbf{Parameter} & \textbf{Value} & \textbf{Reference} \\
    \hline
        \multicolumn{3}{l}{Planet:} \\
        
        $a$ (AU) & $2.513\pm0.00075$ & \citealt{Curiel2011AA...525A..78C} \\
        $e$ & $0.2987\pm0.0072$ & \citealt{Curiel2011AA...525A..78C} \\
        $i$ ($^\circ$)  & $23.758\pm1.316$ & \citealt{McArthur2010ApJ...715.1203M} \\
        $\omega$ ($^\circ$) & $258.82\pm0.43$ & \citealt{Curiel2011AA...525A..78C} \\
        $P$ (days)  &  $1276.46\pm0.57$ & \citealt{Curiel2011AA...525A..78C}  \\ 
        $R_p$ ($R_\mathrm{Jup}$) & 1  & -- \\

        \multicolumn{3}{l}{Ring:} \\
        
        $R_{\rm in}$ ($R_\mathrm{Jup}$) & 1.43 $R_p$ & -- \\
        $R_{\rm out}$ ($R_\mathrm{Jup}$) & 2.47 $R_p$ & -- \\
        $\epsilon$ ($^\circ$) & $30$, $60$, $90$ & -- \\
        $\phi_s$  ($^\circ$)  & $0$ & -- \\
    \hline
    
    \end{tabular}
\end{table}

\section{Discussion} \label{sec:discussion}

The phase curves presented here illustrate how circumplanetary rings can alter the photometry of directly imaged giant planets and affect the interpretation of reflected light observations.
Instruments like Roman, HWO, and the ground-based ELTs will also greatly enhance the sensitivity in high-contrast imaging, requiring a complete understanding of the possible physical phenomena that may exist in order to accurately interpret observations. 
The models presented here offer additional insight on how rings of varying size, obliquity, and orbital geometry affect the reflected light phase curves of giant exoplanets. 
We note that this work focuses solely on the observability of ringed giant planet systems and makes no assumptions about the dynamical stability of the ring configurations explored here.

\subsection{Photometric Signatures of Exorings}

The results presented in Sections~\ref{sec:phase_curves} and~\ref{sec:upsAnd_caseStudy} demonstrate that opaque, icy rings can substantially enhance (or in some geometries, suppress) the reflected light contrast of a giant planet system. 
The magnitude of this effect is strongly dependent on ring size, obliquity, orbital location, and wavelength of the observation. 
At favorable geometries, even a Saturn-equivalent ring around a Jupiter-like planet placed at 5.2\,AU can increase the planet-to-star flux ratio by roughly a factor of 3.1 times in the visible. 
More extended ring systems produce proportionally larger enhancements. 

Beyond the overall affect on the amplitude, the shape of the phase curve is also modified by the ring. 
The ringless planet exhibits the characteristic smooth, bell-shaped phase curve of a Lambertian that peaks near quadrature. 
By contrast, the phase curves in Figures~\ref{fig:ringed_vs_ringless_1-10x}, \ref{fig:phasecurves_30incl_allObl}, and~\ref{fig:phasecurves_color} have varying structure influenced by the ring. 
The $5\times$ and $10\times$ Saturn configurations produce strikingly flat phase curves over the first half of the orbit. 
This flattened shape arises because the ring's reflected flux varies more slowly with orbital phase, and the variation in this model is due to the mutual shadowing by the planet. 
With a larger ring, the shadowing by the planet is reduced to a fractionally smaller area compared to that of a smaller ring, and therefore it has a reduced influence on the overall brightness variation. 

\blue{At certain geometries, rings also induce asymmetries in the phase curve due to mutual shadowing, as can be seen in Figures~\ref{fig:ringed_vs_ringless_1-10x},~\ref{fig:asymmetry}, and~\ref{fig:phasecurves_color}. 
The asymmetries are highlighted in Fig.~\ref{fig:asymmetry} for CGI Band 1. 
An opaque, icy ring would need to be $3\times$ the size of Saturn's rings in order to distinguish the minor asymmetries due to mutual shadowing, while the global asymmetry in the phase curve due to the illuminated/unilluminated ring would be detectable at all sizes.}

Common to all the phase curves modeled in this work is the steep drop in flux at the mid-point of the orbit. 
This transition marks the point in the orbit at which the observer-facing side of the ring plane shifts from the star-illuminated to the unilluminated face. 
Because we model the ring as fully opaque, it contributes reflected light only when the observer is viewing the illuminated face. 
When the ring shifts such that the side facing the observer is no longer illuminated by the host star, it contributes no additional flux and instead partially occults the planetary disk in some cases, suppressing the total system brightness below the ringless case. 
The subtle differences between configurations arise from the varying fractions of shadowing on the planet by the ring, which can be seen in the inset panels of Figure~\ref{fig:ringed_vs_ringless_1-10x}. 

The ring obliquity plays a large role in modulating the shape and amplitude of the phase curve. 
At a system inclination of $i=30^\circ$, the lower obliquities present the illuminated ring face fully to the observer during the first half of the orbit, providing the greatest visible amount of the ring's projected area and reflected flux contribution. 
As the obliquity increases, the projected area of the ring visible to the observer decreases, as does the reflected flux contribution. 
Higher obliquities also increase the degree to which the ring blocks or casts its shadow onto the planetary disk. 
At certain geometries, this can suppress the planetary contribution enough to reduce the total system flux below that of the unobscured ringless planet (e.g., $\epsilon=90^\circ$ in Fig.~\ref{fig:phasecurves_30incl_allObl}). 
The impact of the ring therefore may not be a consistently positive contribution to the flux ratio.

\subsection{Implications for Detection with Roman and Future Observatories}

These exoring phase curve features have direct implications for detectability with Roman CGI. 
The HLC imaging mode in Band 1 is expected to achieve contrasts of $10^{-8}$ or better at angular separations of 3 -- 9 $\lambda/D$. 
A Jupiter-analog at 5.2\,AU peaks in flux at $\sim3.6 \times 10^{-9}$, placing it below the instrument's nominal detection threshold. 
A Saturn-equivalent ring pushes the peak flux ratio to $\sim1.3 \times 10^{-8}$, within striking distance of detectability, while rings of two times Saturn's extent or greater would produce peak signals well above $10^{-8}$, depending on the system geometry. 

Many of the confirmed radial velocity planets that will likely be targeted with Roman orbit at closer separations to their host star than 5.2\,AU. 
For example, 14 Her b (2.839\,AU, $\sim158$\,mas; \citealt{BardalezGagliuffi2025ApJ...988L..18B}), eps Eri b (3.53\,AU, $\leq1103$\,mas; \citealt{Llop-Sayson2021AJ....162..181L}), and HD\,190360 (3.963\,AU, $\leq$247.6\,mas; \citealt{Giovinazzi2025AJ....170...52G}), 
would have estimated peak flux ratios of $1.47\times10^{-8}$, $9.51\times10^{-9}$, and $7.55\times10^{-9}$, respectively (assuming Jupiter-like planets). 
A ring that enhances the contrast by more than 3 times would push the fainter planets into a regime that is more easily detectable with Roman. 
This suggests that reflected light direct imaging missions may preferentially detect ringed giant planets over their ringless counterparts, if they are readily present in the population. 
An example target list for Roman with the potential flux ratio enhancement provided by a ring is provided in \textit{Lane et al., (submitted)}.

However, opaque rings do not uniformly enhance the system brightness at all geometries. 
In the worst case scenario, a ringed planet could appear fainter than a ringless planet of the same size and atmospheric composition. 
This is a counterintuitive result with important consequences for planning and interpreting multi-epoch observations. 

The sensitivity of the phase curve to the ring geometry highlights the value of multi-epoch observations to sample the phase curve and distinguish ring-induced variability from the phase behavior of a ringless planet. 
The sharp flux transition at the orbital midpoint, where the ring shifts from the illuminated to unilluminated face, is a distinctive feature absent in ringless phase curves and could serve as a strong discriminant when it falls within the observing windows.
However, we note that this feature is unique to fully opaque rings and will likely not persist for rings of lower opacity, which will instead have multi-peaked phase curves (\citealt{Arnold2004-gc, Dyudina2005-zy, Veenstra2025A&A...693A.310V}; \textit{Lane et al., submitted}). 

As shown in Figure~\ref{fig:phasecurves_color}, icy rings maintain high reflectivity across visible-wavelengths, including those covering the methane absorption dominated regions of the giant planet's spectrum (e.g., Bands 4 and 3C), where the planet's flux is substantially reduced. 
A ringed giant planet observed in a methane-band filter would appear anomalously bright relative to expectations for a ringless giant planet atmosphere. 
A color comparison may serve as a discriminant for ring detection, even when the brightness enhancement provided by the ring is only moderate (for example, see Fig.~\ref{fig:color_comparison}). 
\blue{A photometric uncertainty of better than $1{\rm e-}9$ would be sufficient to distinguish between ringed and ringless cases in color space.}

\blue{Additionally, the distinct spectral features of icy rings could aid in their identification in near-infrared spectra. 
While Saturn's ring spectra are relatively featureless in the optical, observations of the main rings (A, B, C) show that their spectra between $\sim1$--$2.5\,\mu$m are dominated by water ice and exhibit prominent water-ice absorption bands \citep{Nicholson2008Icar..193..182N}.
Such distinct features could provide a potential diagnostic of rings around widely separated giant planets, where their atmospheres are not expected to show strong water-ice absorption features at these wavelengths. 
For example, the spectra of Jupiter and Saturn do not exhibit strong water features because the water is trapped deep in their atmospheres \citep{Encrenaz1999A&ARv...9..171E, Nicholson2008Icar..193..182N}. 
In addition, changes in the strength of ring-related absorption features with orbital phase may provide another means of distinguishing ring signatures from a planet's atmospheric constituents, particularly if their phase dependence differs from that expected for atmospheric species.}

The obliquity-dependence of the flux ratio is also relevant for scheduling and interpreting observations of planets like $\upsilon$ And d. 
Due to the orbital geometry of the planet, $\upsilon$ And d is accessible to HLC into early 2027, before falling behind the instrument's inner working angle for about one year. 
These windows offer the opportunity to sample the phase curve of the planet, and may have optimistic regions for detection if a ring is present to enhance the signal with a favorable obliquity similar to the range tested here.
If the obliquity is instead closer to edge-on, this same window offers the less favorable circumstance for detection. 
The sharp discontinuities in flux in the left and center panels of Figure~\ref{fig:upsAnd_phasecurves} mark the points in the orbit where the ring transitions from its illuminated to unilluminated face with respect to the observer. 
This transition region indicates an additional distinct, opaque-ring-specific signature that would be absent from the phase curve of a ringless planet. 

The region where the ring shifts from its lit to unlit side should thus be considered when scheduling multi-epoch observations. 
Observers should look for abrupt changes in brightness between epochs that could indicate a transition between the illuminated and unilluminated faces of an opaque ring. 
Sampling observations near or around these transitions, where feasible, could provide a distinctive diagnostic of rings that would not be expected from a ringless planet with only photometric observations due to the extreme flux variation.

\subsection{Implications for Planetary Characterization}

If ring contributions are not accounted for in the interpretation of direct images, they may introduce biases in the inferred planetary properties. 
An unrecognized ring could cause the planet to appear brighter (or dimmer) than expected, leading to an incorrect estimate of the geometric albedo or planetary radius. 
\blue{For a planet with a Saturn-like opaque ring, if one assumed a planetary radius of $1\,R_{\rm Jup}$, an albedo greater than one would be required to explain the measured photometry at quadrature.
If an observer instead assumes a Jupiter-like albedo (B1), a planetary radius of $2.46\,R_{\rm Jup}$ would be required to explain the measurement. 
If the ring is face-on with respect to the observer, a planetary radius of $2.66\,R_{\rm Jup}$ would be required to explain the photometry.
Therefore, a Saturn-sized ring might lead us to incorrectly assume that a planet has a radius that is inflated by 2.5 times or more. 
A larger ring of $2\times$ the size of Saturn's would falsely lead us to believe that the planet has a radius more than $4\times$ that of Jupiter.}

In the most problematic scenario, a ringed planet observed sparsely across multiple epochs could produce brightness variations that may be incorrectly inferred as a second point source in the system, particularly if the point source appears at different projected separation across observations. 
In such a scenario, an incorrect assumption of additional point sources in the system could introduce confusion \citep{Keithly2021ApJ...919L..11K, Pogorelyuk2022ApJ...937...66P, Hasler2026JATIS..12d1008H}. 
Resolving such ambiguities may ultimately require spectral characterization, polarimetric observations \citep{Lietzow2023A&A...671A.113L, Veenstra2025A&A...693A.310V}, or additional monitoring with sufficient sampling to gather the shape of the phase curve.

\blue{As illustrated in Figure~\ref{fig:planet_rings_filters}, rings may also alter the apparent strength of atmospheric absorption features. 
Neglecting the ring contribution to the combined planet-ring spectrum could lead to underestimates of the abundances of atmospheric constituents. 
For example, shallower methane absorption bands in the ringed planet spectrum could be misinterpreted as evidence for lower atmospheric methane abundance.}

Additionally, a ring in an unfavorable orientation that obstructs the planet disk and is unlit with respect to the observer would decrease the total system flux ratio below that of the predicted ringless planet, making targets unreachable below the detection limit.
While there are many unknowns in observation planning for reflected light planets (e.g., atmospheric properties, priors on orbital parameters for RV-only planets), exorings present another influencing factor to our flux ratio estimates that should be considered in planning and interpretation. 
Accounting for circumplanetary structures in our models will be a necessary step as we push to imaging mature planets at high-contrasts, where ring systems may be found. 

\subsection{Comparison to Previous Models and Future Work}\label{sec:future_work}

The model presented here offers a method for estimating the first-order detectability of exoplanets with opaque, icy rings in direct images and will be be expanded in the future. 
Previous numerical models have also demonstrated that exorings can alter the phase curves of exoplanets. 
In particular, \cite{Zuluaga2022ascl.soft05016Z} and \cite{Veenstra2025A&A...693A.310V} developed the \texttt{Pryngles} code to calculate the light curves of transiting and non-transiting exoplanets. 
Their Python framework uses a surface discretization scheme coupled with radiative-transfer calculations and laboratory measurements of olivine to model the ring-particle scattering. 
The \texttt{Pryngles} model accounts for ring reflection, transmission, mutual-shadowing, and ring polarization. 

Our model takes a complementary approach focused on directly imaged giant planets. 
Rather than modeling the scattering properties of individual ring particles, we use the empirically measured spectra of Saturn's rings to represent the wavelength-dependent ring reflectance, coupled to a ray-casting framework that determines the illuminated and visible portions of the planet and ring and accounts for mutual shadowing and occultation. 
This allows us to efficiently explore the effects of ring size, obliquity, orbital inclination, and orbital phase on the observable flux ratios of ringed planets. 
In contrast to the partially transmitting rings considered by \cite{Veenstra2025A&A...693A.310V}, our model currently assumes fully opaque, icy rings, providing a limiting case to be improved in future model expansions. 

Real ring systems span a range of optical depths and scattering properties. 
While regions of Saturn's ring system are optically thick \citep{Tiscareno2019Sci...364.1017T, Miller2024SSRv..220...70M}, partially transparent rings could contribute varying amounts of reflected flux and true ring particles will have more complex scattering behavior that will influence the full range of the phase curve. 
The ring reflectance in our model is fixed to a single empirically measured spectrum of Saturn's B ring, which is dominated by water ice. 
This provides a well-characterized and physically-motivated baseline, but will be extended to capture a broader range of potential ring constituents that may exist around exoplanets in future updates.
We will also expand the ring model to include a range of optical depths and ring compositions beyond the opaque, water-ice baseline presented here, as these properties significantly affect both the reflected flux and mutual shadowing components.
Additionally, extending the forward model to couple with retrieval codes would provide an avenue for direct interpretation of observations and could provide critical guidance for observation planning with Roman and HWO. 

\section{Summary} \label{sec:summary}

In this work, we present a  ringed planet model that combines a ray casting framework with empirical planet and ring reflectance spectra to simulate the reflected light phase curves of giant planets with opaque, icy ring systems as a function of ring size, obliquity, and orbital geometry\footnote{Model available on GitHub (\href{https://github.com/snhasler/exorings-hci}{https://github.com/snhasler/exorings-hci}) \blue{and Zenodo \citep{exorings_zenodo}}}. 
We apply the model to Jupiter-like planets with ring extents from $1 - 10\times$ Saturn's A, B, and C ring radii, and to the known radial velocity target $\upsilon$ Andromedae d as a test case for Roman CGI observations. 
Our key findings are: 

\begin{itemize}
    \item Opaque, icy rings can increase the reflected light flux of a giant planet system at favorable geometries by $3.1\times$ to more than an order of magnitude. 

    \item Photometric variation induced by rings is sensitive to orbital location, ring obliquity, and viewing geometry. 
    Rings do not uniformly brighten the system and may obstruct or shadow the planet disk at certain geometries, reducing the total system flux ratio below that of an equivalent ringless planet. 

    \item Distinctly identifying the presence of a ring will likely require multi-epoch observations that sample the shape of the phase curve. 

    \item Multi-band observations leveraging the flux ratio between methane-band and wide bandpass filters may provide a complementary color diagnostic for exoring detection. This is in addition to polarization signals, which we do not discuss but have been explored in the literature (e.g., \citealt{Lietzow2023A&A...671A.113L, Veenstra2025A&A...693A.310V}) and will be possible with Roman. 

    \item A Saturn-equivalent ring around the known radial-velocity planet $\upsilon$ And d can enhance the peak system flux by up to $4.2\times$ for a favorable ring obliquity ($\epsilon=30^\circ$), pushing it above an SNR of 5 threshold for detection with Roman; an edge-on ring ($\epsilon=90^\circ$) suppresses the flux ratio below the ringless case.

    \item Ring contributions may bias inferred planetary properties if not included in the forward models used to interpret reflected light observations.

\end{itemize}

\begin{acknowledgments}
\blue{We thank Tyler D.~Robinson for the thoughtful and constructive feedback, which improved the clarity and scope of the manuscript.}
We would like to thank Joshua Colwell and Robert West for insightful discussion regarding the rings of Saturn. 
S.N.H.\ would also like to thank Karl Stapelfeldt for many thoughtful discussions on reflected light direct imaging.
This research was carried out as part of the IPAC at California Institute of Technology Visiting Graduate Student Fellowship and at the Massachusetts Institute of Technology. A portion of the work was carried out at the Jet Propulsion Laboratory, California Institute of Technology, under a contract with the National Aeronautics and Space Administration (80NM0018D0004). This work was co-authored by employees of Caltech/IPAC under Contract No.~80GSFC21R0032 with the National Aeronautics and Space Administration.
This research has made use of the NASA Exoplanet Archive, which is operated by the California Institute of Technology, under contract with the National Aeronautics and Space Administration under the Exoplanet Exploration Program.
This research has made use of the SIMBAD database, operated at CDS, Strasbourg, France.
\end{acknowledgments}

\vspace{5mm}

% \begin{contribution}

% List contributions.

%% But authors are expected to provide more specific details, e.g. 
%%
%%SC was responsible for writing and submitting the manuscript.
%%WWM came up with the initial research concept and edited the manuscript.
%%OTS obtained the funding and edited the manuscript.
%%EBF provided the formal analysis and validation. He also edited the manuscript.
%%GEH Supervised the undergraduates, wrote the software and administers the project github and Zenodo repositories.
%%
%% Authors can use the Contributor Role Taxonomy (CRediT) at
%% for ideas on how write a good statement tailored to their needs.

% \end{contribution}

\software{\texttt{Astropy} \citep{2013A&A...558A..33A,2018AJ....156..123A}, \texttt{SciPy} \citep{2020SciPy-NMeth}, \texttt{NumPy} \citep{harris2020array}, \texttt{Matplotlib} \citep{Hunter:2007}, \texttt{corgietc} \citep{corgietc}, \texttt{pandas} \citep{reback2020pandas}
}

\bibliography{references}{}
\bibliographystyle{aasjournalv7}

\end{document}